\documentstyle[11pt,epsf,subfigure]{article}

\newcommand{\rf}[1]{(\ref{#1})}
\def\void{}
\def\labelmark{}

\newenvironment{formula}[1]{\def\labelname{#1}
\ifx\void\labelname\def\junk{\begin{displaymath}}
\else\def\junk{\begin{equation}\label{\labelname}}\fi\junk}%
{\ifx\void\labelname\def\junk{\end{displaymath}}
\else\def\junk{\end{equation}}\fi\junk\labelmark\def\labelname{}}

{\ifx\void\labelname\def\junk{\end{array}\end{displaymath}}
\else\def\junk{\end{array}\right.\end{equation}}
\fi\junk\labelmark\def\labelname{}\def\junk{}
}

\newcommand{\beq}{\begin{formula}}
\newcommand{\eeq}{\end{formula}}
\newcommand{\beqv}{\begin{formula}{}}
\newcommand{\bea}{\begin{eqnarray}}
\newcommand{\eea}{\end{eqnarray}}

\begin{document}

\begin{titlepage}

\null
\begin{flushright}
SU-4240-607
\end{flushright}
\vspace{20mm}

\begin{center}
\bf\Large Scaling and the Fractal Geometry\\
          of Two-Dimensional Quantum Gravity
\end{center}

\vspace{5mm}

\begin{center}
{\bf S. Catterall}\\
{\bf G. Thorleifsson}\\
{\bf M. Bowick}\\
{\bf V. John}\\
\vspace{2mm}
Physics Department, Syracuse University,\\
Syracuse, NY 13244.
\end{center}

\begin{center}
\today
\end{center}

\vspace{10mm}
      
\begin{abstract}
We examine the scaling of geodesic correlation functions
in two-dimensional gravity and in spin systems coupled
to gravity. The numerical
data support the scaling hypothesis and indicate that
the quantum geometry develops a non-perturbative
length scale.
The existence of this length scale
allows us to extract a Hausdorff dimension.
In the case of pure gravity we find $d_H \approx 3.8$,
in support of recent theoretical calculations that $d_H = 4$. 
We also discuss the back-reaction of matter on the geometry.
\end{abstract}
\vfill

\end{titlepage}

\section{Introduction}

Remarkable strides have been made in recent years in our understanding
of the properties of two-dimensional quantum gravity \cite{rev}.
Calculations carried out
within the framework of conformal field theory have yielded
the gravitational dressing of {\it integrated} matter field operators,
correlation functions on the sphere and the torus partition
function. On the other hand matrix models have provided us with a 
powerful calculational tool that enables us to compute the above mentioned
quantities and also to perform  
the non-perturbative sum over topologies. 

Nevertheless there are still important geometrical quantities
of physical interest that are not well understood analytically. 
Perhaps the most fundamental is the intrinsic Hausdorff dimension of the
typical surface generated by the coupling of 
matter to $2d$-gravity \cite{dav,wat}.   
One may think of the Hausdorff dimension as an order parameter 
characterizing possible phases of the theory.
If there exists a power law relation between two reparametrization
invariant quantities with the dimension of volume ($V$) and length ($L$),
this provides a well-defined fractal dimension ($d_H$) via 
$V \propto L^{d_H}$.
As there is no natural notion of a length scale in these theories, 
one has to be introduced by hand, at least in the
continuum formulation. 
In the discretized approach this length scale is
provided by the short distance cut-off 
corresponding to the finite elementary link length.

Recently a transfer matrix formalism utilizing matrix model
amplitudes has been developed that predicts
the Hausdorff dimension $d_H = 4$ for
pure $2d$ gravity \cite{KKMW}. This approach has not
yet been extended to the case of unitary minimal 
models coupled to gravity.  On the other hand
the analysis of the diffusion equation for a random
walk on the ensemble of $2d$ manifolds determined by
the Liouville action yields a prediction for the 
Hausdorff dimension which agrees with the transfer matrix
approach for pure gravity. It may also be extended to
include the coupling of conformal matter of central
charge $c \leq 1$ \cite{moto}. 

These analytic predictions
for the Hausdorff dimension rely on the validity of
certain scaling assumptions.  It also appears that
there are several potentially inequivalent definitions
of an appropriate fractal dimensionality.
It seems very worthwhile therefore to explore these
issues numerically.   
Earlier numerical work addressing this question has been
remarkably inconclusive \cite{billoire,mig,burda}. 
Indeed for a while it was claimed that there was 
no well-defined Hausdorff dimension in the
case of pure gravity \cite{mig}. In contrast clear numerical
evidence for a fractal scaling of gravity coupled
to $c=-2$ matter was found in \cite{KKSW}.

In this letter we establish numerically that this scaling
behavior is valid for pure gravity as well as the Ising
and 3-state Potts models coupled to gravity. 
We employ a careful finite size
scaling analysis of appropriate correlation
functions.  For pure gravity we find $d_H \approx
3.8$ in qualitative agreement with \cite{wat,KKMW,moto}. 
For the Ising and 3-state Potts models the 
values of $d_H$ that we obtain do not seem to
detect the back reaction of matter on the geometry. 

This paper is organized as follows.
In section 2 we describe the application of finite
size scaling to loop-loop correlation functions.
In section 3 we outline our numerical procedures
and results. In section 4 we present the existing
theoretical predictions for the Hausdorff dimension. 
Finally section 5 is a discussion of our conclusions.

\section{Scaling} 

Finite size scaling is a well-established technique for
analyzing the critical behavior of conventional statistical
mechanical models \cite{barber}. In numerical studies of
quantum gravity it has traditionally been employed in
a rather limited context - typically by extracting a power
law scaling for integrated matter field operators at the
critical point.

In general, the scaling hypothesis asserts that near a critical point 
an observable $O$, a function of two variables $x$ and $y$,
will depend on only one
{\it scaling} combination $\mu=y/x^q$ up to an overall power factor
$x^p$
\beq{*2}
O\left(x,y\right)\sim x^pf\left(y/x^q\right).
\eeq
The powers $p$ and $q$ are related to the critical
exponents of the model.
We will test this hypothesis by analyzing geodesic correlators
defined on dynamical triangulations. 

The fundamental objects in two-dimensional gravity are loop-loop
correlators. To define these consider two marked loops of
length $l$ and $l^\prime$ on a triangulation. 
If we define a geodesic
distance $r$ between the loops on the graph as the minimal
number of links that must be traversed to go from $l$ to $l^\prime$,
we can define a correlation function 
$n_{l,l^\prime}\left(r\right)$ as 
\beq{*201}
n_{l,l^\prime}(r,N)=
\sum_{T\in{\cal T}_2(N)} 1 \;.
\eeq
In this expression ${\cal T}_2(N)$ refers to the class of
triangulations with $N$ triangles and two loops of
length $l$ and $l^\prime$ separated by a geodesic distance $r$. 
As defined above $n_{l,l^\prime}(r,N)$ is proportional to
the number of triangulations satisfying the above constraints. 
We chose to work in the microcanonical ensemble as it
is convenient computationally  
and the effect of restricting
to fixed volume can be exploited in the
finite size scaling analysis.  
The point-point correlator 
$n(r,N)$, which counts the number of triangulations
with two marked points separated by geodesic
distance $r$, can now be
obtained from Eq.\ (\ref{*201}) in the limit that the
lengths of the loops are taken to zero.

The scaling hypothesis applied to $n\left(r,N\right)$ implies
\beq{*scal1}
n\left(r,N\right)=N^p\;n\left(r/N^q\right) \;.
\eeq
The combination $l_G=N^q$ constitutes a dynamical length scale which
appears {\it non-perturbatively} in the theory. It can be used
to define a Hausdorff or fractal dimension $d_H=1/q$ characterizing
the quantum geometry.
Notice that in this case the exponent $p$ is not free ${-}$ it is
constrained by the fact that the integral of $n\left(r,N\right)$
over all geodesic distances recovers the total number of
points $N$. This yields $p=1-1/d_H$.
It is easy to measure the point-point correlator
numerically and thus determine $d_H$.

This discussion can be generalized to include spin models
coupled to gravity.  In this case the boundary loops
will be dressed with fixed boundary spin configurations.
For the point-point correlator we can use 
the symmetry of the spin models to  reduce the possible correlators 
to two distinct types, which we denote $f_1\left(r, N\right)$ and
$f_2\left(r, N\right)$. The correlator
$f_1$ counts the number of points
at distance $r$ for which the spin variable is in the same state as
the initial marked point. The correlator $f_2$ counts the number
of spins in a different state from the initial marked point. 
The total number of points at geodesic distance $r$ is then
\beq{*202}
n\left(r, N\right)=f_1\left(r,N\right)+f_2\left(r,N\right).
\eeq 
The scaling hypothesis can be applied as before to these correlators,
resulting in a definition of $d_H$.

To define spin-spin correlators we note that the spin 
variables of the $q$-state
Potts model may be taken to be the unit vectors $\vec{e}_\alpha$ of
a hyper-tetrahedron in  $q-1$ dimensional space. We can then
identify the product of two spins as the scalar product of the associated
link vectors
\beq{*111}
\sigma_\alpha \sigma_\beta \equiv \vec{e}_\alpha \cdot \vec{e}_\beta 
= \left( 1 + \frac{1}{q-1}\right)\delta_{\alpha \beta} - \frac{1}{q-1}.
\eeq
The (unnormalized) spin-spin correlator with one marked point $i$ is 
\beq{*112}
g_{un}(r,N) = \sum_{T} \sum_{\{\sigma_i\}} 
\sum_j \sigma_i \sigma_j \; \delta(d_{ij} - r) 
\; e^{{\textstyle -S_N(\sigma ,T)}}\;\;.
\eeq
In terms of the distributions $f_1$ and $f_2$
defined earlier we see from Eq.\ (\ref{*111}) that this may be written as
\beq{*203}
g_{un}(r,N)=f_1\left(r, N\right)-{1\over
q-1}f_2\left(r, N\right).
\eeq
The scaling hypothesis for $g_{un}$ takes the form
\beq{*204}
g_{un}(r,N) = N^{\frac{\gamma}{\nu d_H} -s} g_{un}(r/N^s).
\eeq
The overall power is again determined from the constraint that the integral
of $g_{un}(r,N)$ is just the usual 
magnetic susceptibility, which scales at criticality 
as $\chi\sim N^{{\gamma\over\nu d_H}}$.
If we make the assumption that these critical systems contain only
one length scale then it is natural to assume that both
$n(r,N)$ and $g_{un}(r,N)$ depend on the same scaling variable. 
This implies that $s=1/d_H$, where the Hausdorff dimension $d_H$ is now
that appropriate to the matter-coupled theory. We shall re-examine
this assumption critically in light of the numerical results in
the final section of the paper. 
 
We will also consider the normalized spin-spin correlator with one marked point $i$: 
\beq{*113}
g_{n}(r,N) = \sum_{T} \frac{1}{\sum_j \delta(d_{ij} -r)} 
\sum_{\{\sigma_i\}} \sum_j \sigma_i \sigma_j \;\delta(d_{ij} - r) 
\; e^{{\textstyle -S_N(\sigma ,T)}}\;\;.
\eeq

\section{Numerical Simulations}

To investigate the validity of the scaling hypothesis
we have performed Monte Carlo simulations on three models;
pure gravity (central charge $c=0$), the Ising model $(c=1/2)$ and
3-state Potts model $(c=4/5)$ coupled to gravity.
In the microcanonical ensemble the partition function of 
these models is given by
\beq{*21}
Z(\beta,N) = \sum_{T\in {\cal T}} Z_M(\beta,N)
\eeq
where $Z_M(N,\beta)$ describes the matter sector (absent
for pure gravity). For a $q$-state Potts model this is
\beq{*22}
Z_M(\beta,N) = \sum_{\{\sigma_i\}} {\rm exp} \left ( \beta \sum_{<i,j>}
(\delta_{\sigma_i,\sigma_j} - 1) \right ),
\eeq
were $\sigma_i$ denote the Potts spins, $i$ denotes a
lattice site and $<i,j>$ indicates that the sum is over nearest-
neighbor pairs on the lattice.  

The integration over manifolds is implemented
as a sum over an appropriate class of triangulations ${\cal T}$.
Since it has been observed that finite size effects in
numerical determinations of critical exponents are
generally smaller if one includes degenerate triangulations
in ${\cal T}$, i.e. triangulations allowing two vertices
connected by more than one link and vertices connected to
itself \cite{gud}\footnote{This corresponds to allowing tadpoles and
self-energy diagrams in the dual lattice formulation.},
we will work in that ensemble.

In the simulations a standard link-flip algorithm
was used to explore the space of triangulations and a
Swendsen-Wang cluster algorithm employed for the spin updates.
Lattice sizes ranging from 500 to 32000 triangles
were studied and typically $10^6$ to $4\times 10^6$
Monte Carlo sweeps performed for each lattice size
(a sweep consists of flipping about $N$ links and one 
SW update of the spin configuration).

\subsection{Pure Gravity}

We start with the results for pure gravity.  Here we 
measured the point-point distributions $n(r,N)$ 
both on the direct and the dual lattice. On the dual lattice
geodesic distances are measured as shortest paths going from
one vertex to another. 
Having measured these distributions for different lattice sizes
there are several ways
we can use the scaling assumption Eq.\ \rf{*scal1} to extract $d_H$.
We use two methods. 

First we fitted a distribution (for a given lattice size) 
to an appropriately chosen function from which we located
the maximum of the distribution $r_0$ and its maximal
value $n(r_0)$. Then the scaling assumption implies that
$r_0 \sim N^{1/d_H}$ and $n(r_0) \sim N^{1-1/d_H}$.
We fit to the function 
\beq{*23}
P_l(r) \:{\rm exp}(-ar^b) \;. 
\eeq
The exponential is included in order to capture the 
long-distance behavior of the distribution and $P_l$ is
an $l$-order polynomial.  The
order of the polynomial is chosen in such way that
we get a reasonably good fit; a
4{\it th} order polynomial turned out to be sufficient.  
We checked that
the values of $r_0$ and $n(r_0)$
did not change appreciably if we increased the order of $P_l(r)$. 
The values of $r_0$ and $n(r_0)$ obtained in this way are plotted 
in Figs.\ 1{\it a} and 1{\it b} on log-log plots.  
As expected both
quantities scale well with $N$ (significantly better for the direct 
lattice). The Hausdorff dimensions extracted from the
slopes are listed in Table 1.

\begin{figure}
\begin{center}
\subfigure{
\epsfxsize=2.6in \epsfbox{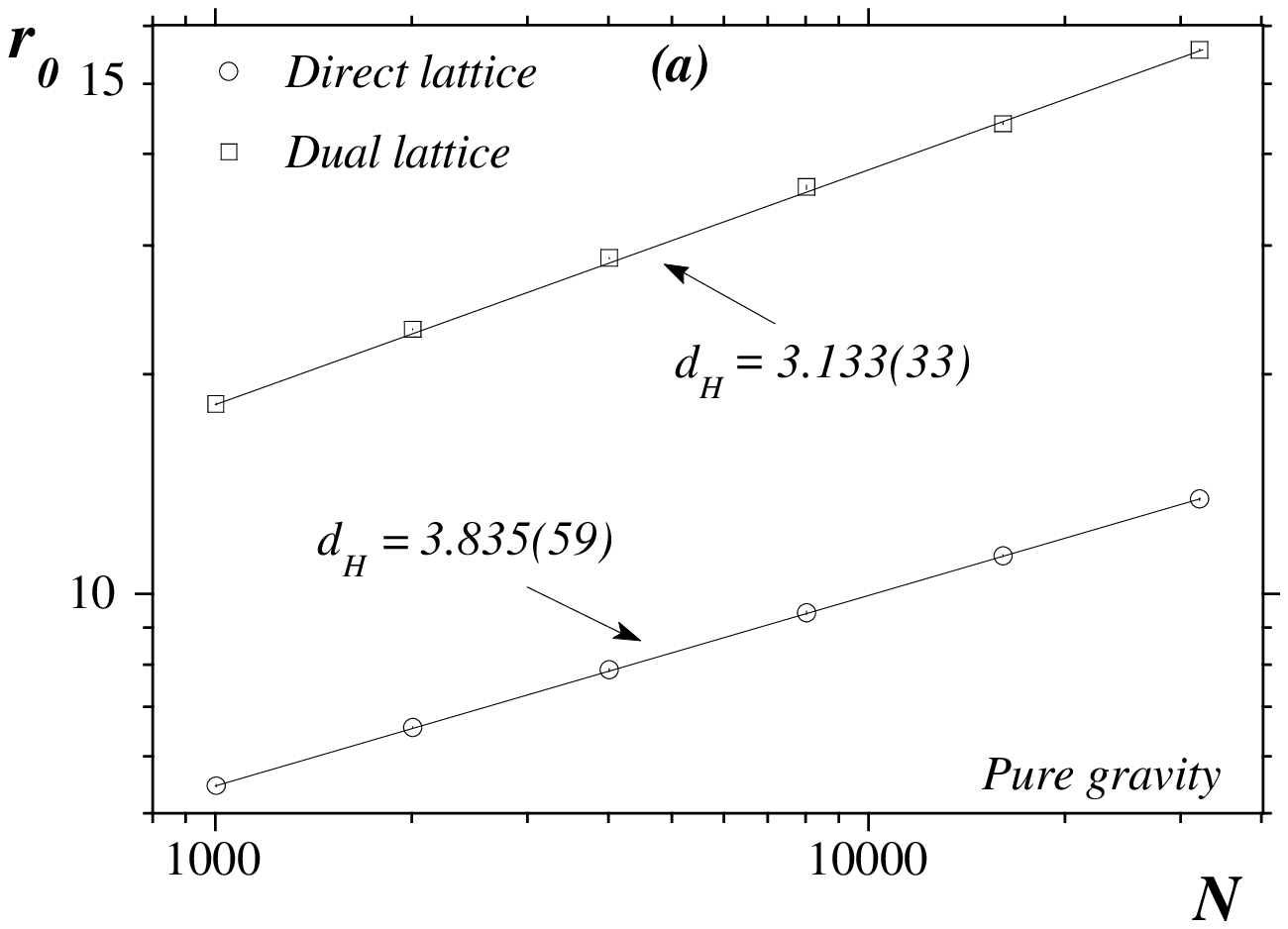} }
\subfigure{
\epsfxsize=2.6in \epsfbox{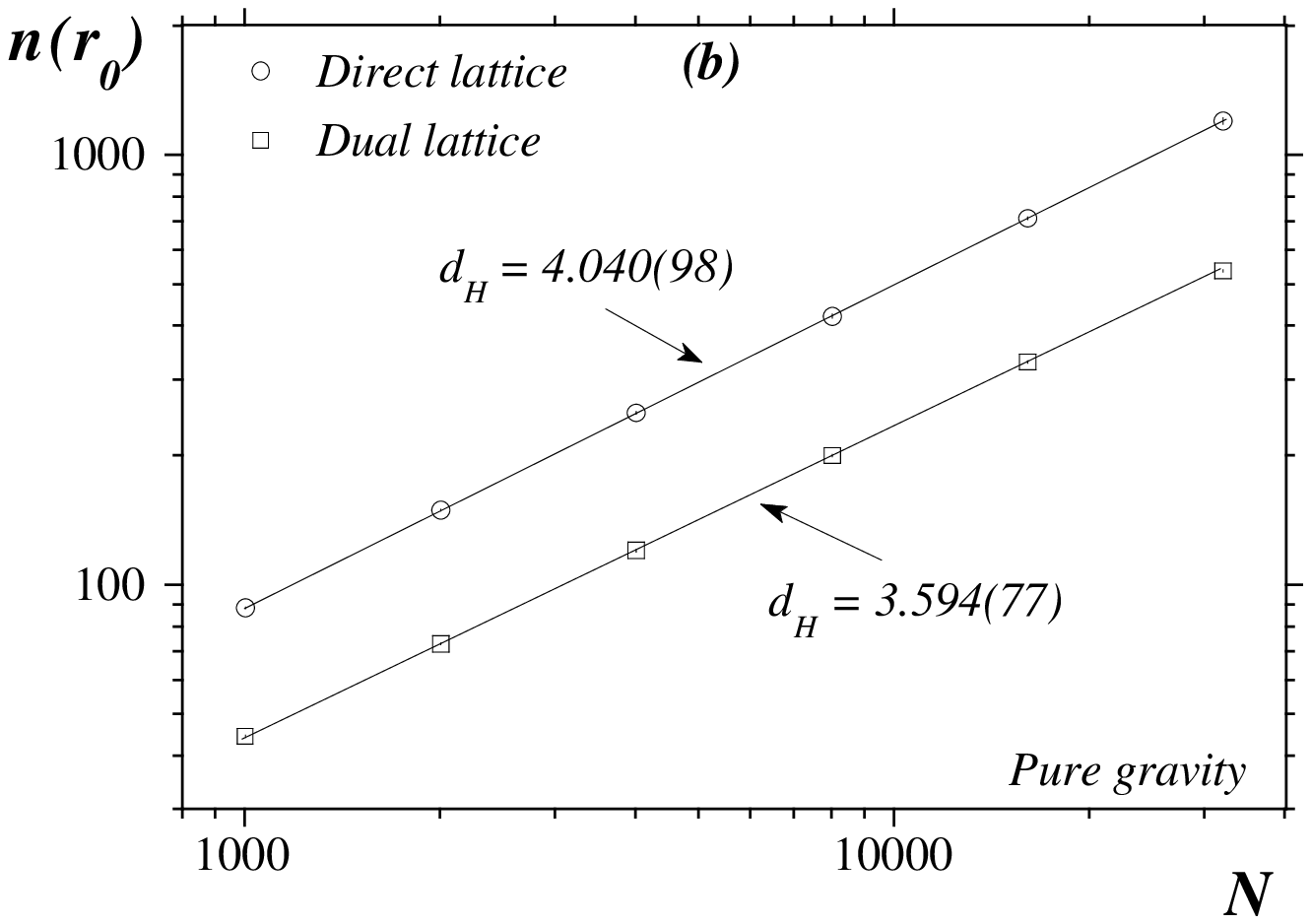} }
\end{center}
\caption
 { Volume scaling of (${\it a}$) the location of the 
   peak $r_0$ in the distributions $n(r,N)$ 
   and ({\it b}) their maximal value $n(r_0)$
   in the case of pure gravity.  Data is shown both
   for the direct and dual lattices and the extracted values
   of $d_H$ are included.}
\label{fig:1}
\end{figure}

{\small
\begin{table}
\begin{center}
\begin{tabular}{|l|cl|cl|}  \hline
  & \multicolumn{2}{c|}{Direct lattice}
  & \multicolumn{2}{c|}{Dual lattice}   \\
  &  $d_H$  & $\tilde{\chi}^2$  &  $d_H$  &  
     $\tilde{\chi}^2$  \\ \hline
{\it (a)} &&&& \\
   $126-250$     &  3.640(60)  &  44.6  & 
                    2.497(37)  &  49.2    \\
   $250-500$     &  3.707(45)  &  13.0  & 
                    2.715(40)  &  29.1    \\ 
   $500-1000$    &  3.727(42)  &  8.0   & 
                    2.871(38)  &  20.5    \\
   $1000-2000$   &  3.770(38)  &  4.2   & 
                    2.996(26)  &  22.6    \\
   $2000-4000$   &  3.800(54)  &  2.3   & 
                    3.111(39)  &  12.5    \\
   $4000-8000$   &  3.804(55)  &  1.5   & 
                    3.217(47)  &  9.7     \\
   $8000-16000$  &  3.810(55)  &  0.97  &
                    3.264(34)  &  6.9     \\
   $16000-32000$ &  3.830(50)  &  1.4   & 
                     3.411(89)  &  4.8     \\  \hline
{\it (b)} &&&& \\
   $1000-32000$  &  3.790(30)  &  13.0  &
                     3.150(31)  &  85    \\  \hline
{\it (c)} &&&& \\
   position &  $3.835(59)$  & 0.03   & $3.133(43) $ & 10.45  \\
   height   &  $4.040(98)$  & 0.09   & $3.594(77) $ & 0.37   \\ \hline
\end{tabular}
\end{center}
\caption{Extracted values of $d_H$ from $n(r,N)$ in the case of
pure gravity.
The values in {\it (a)} are obtained
by collapsing data for two consecutive 
lattices sizes on a single curve using
{\it one} scaling parameter.  {\it (b)} is the same except data from
all lattice sizes between 1000 and 32000 triangles are used.
In {\it (c)} the values are obtained from the volume scaling
of $r_0$ and $n(r_0)$ separately.
The quality of the fit is indicated by $\tilde{\chi}^2$ 
and the
errors in {\it (a)} and {\it (b)} indicate where
$\tilde{\chi}^2$ changes by one unit from its minimal value.}
\end{table}

}

Another way to extract $d_H$ is to use the scaling
relations directly to collapse distributions
for different lattices sizes on to the same curve, 
using only a {\it single} scaling parameter $d_H$.
This we have done including all the data (for $N \geq 1000$)
and also, to explore the finite size corrections, 
using only pairs of datasets ($N$ and $2N$).
The same functional form 
Eq.\ (\ref{*23}) was used in the fits.  The results are shown in Table 1, 
together with the quality of the fits 
($\tilde{\chi}^2 = \chi^2 / {\it dof}$).
The errors quoted indicate where
$\tilde{\chi}^2$ changes by one unit from its minimal value. 
In Figs.\ 2{\it a} and 2{\it b} we show overall scaling plots for
$n(r,N)$, both for the direct and dual lattices.

\begin{figure}
\begin{center}
\subfigure{
\epsfxsize=2.6in \epsfbox{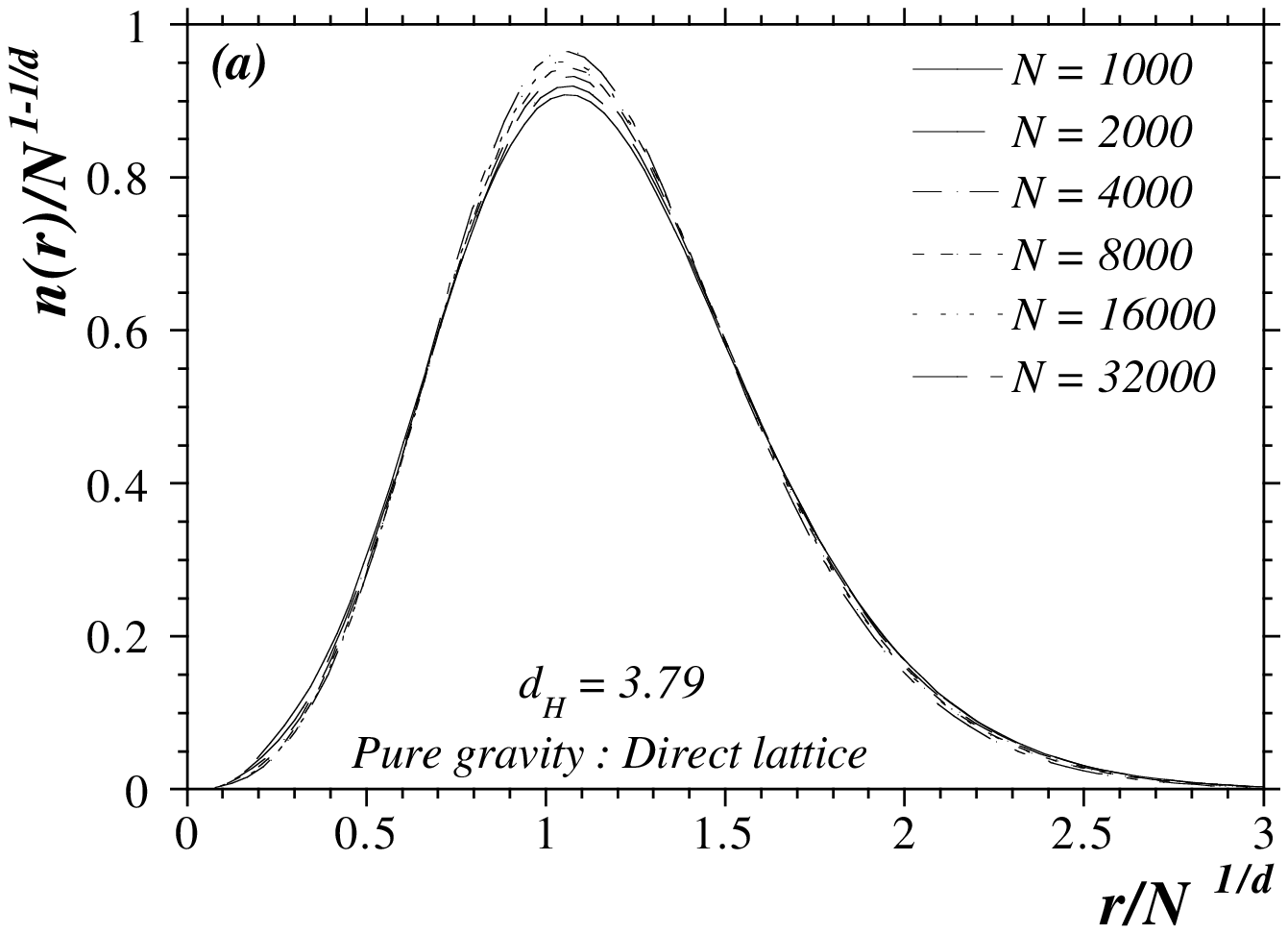} }
\subfigure{
\epsfxsize=2.6in \epsfbox{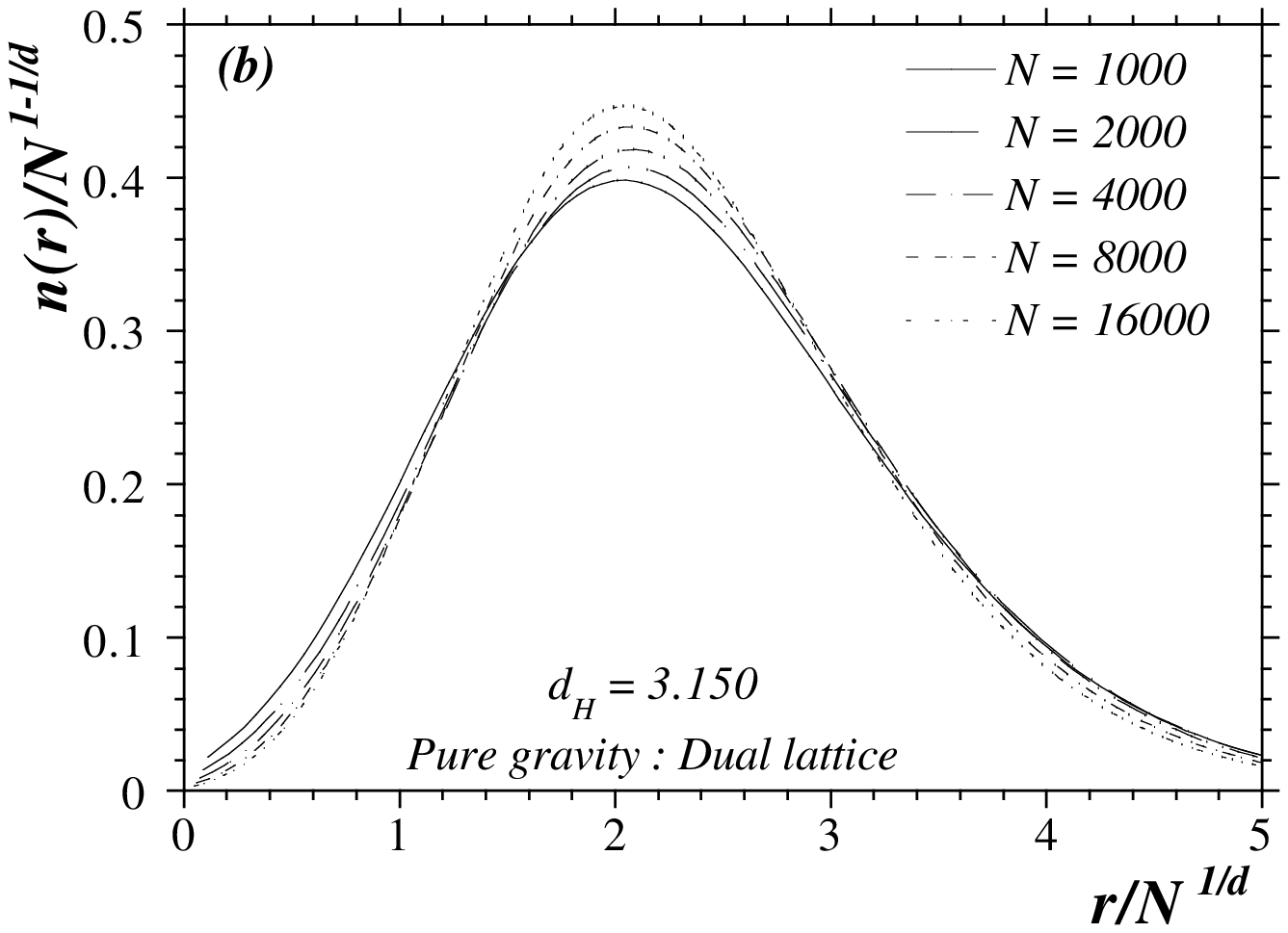} }
\end{center}
\caption
 { Scaling plots for the point-point distributions $n(r,N)$ in the
   case of pure gravity; {\it (a)} the direct and
   {\it (b)} dual lattice. Shown are the curves fitted to distributions 
   after rescaling. 
   The value of $d_H$ is chosen so as it minimized
   the total chi-square of the fits.} 
\label{fig:2}
\end{figure}

From these results we can immediately draw a number
of conclusions. Consider first
the direct lattice.  Fig.\ 2{\it a} shows 
that the scaling hypothesis is indeed well satisfied for 
the distribution $n(r,N)$. This is also evident from the low
values of $\tilde{\chi}^2$ for the fits (Table 1). 
The values of $d_H$ obtained from the scaling of $r_0$ and
$n(r_0)$ and also from collapsing the data are close to
the expected value of $d_H = 4$.  These results
are obtained on moderately small lattices, illustrating the 
superiority of this method of extracting $d_H$ to earlier
numerical attempts.

But we also notice that there is a systematic
increase in the value of $d_H$ with lattice size.  Even though
this effect is too small compared to the uncertainty in 
the measured values to allow reliable 
extrapolation to infinite volume
$d_H$, it indicates that the difference
between measured and expected values of $d_H$ is
due to finite-size effects.  
The improvement of the $\tilde{\chi}^2$ values of
the fits with increasing lattice size also
implies diminishing deviations from scaling. 

It is also intriguing that the scaling of the peak
heights seems to give better values of $d_H$ (close to
the theoretical results for the direct lattice).
Since the heights of the peaks take continuous
values, as opposed to the discrete geodesic
distance, it plausible that they are less sensitive
to the discretization

On the dual lattice we observe 
much larger finite size deviations.
This is evident both from
Fig.\ 2{\it b} and the values of $\tilde{\chi}^2$ in Table 1.
This is not hard to understand.  The short distance behavior
of $n(r,N)$ is dominated by a power 
growth $r^{d_H-1}$.  But as the order of vertices
on the dual lattice is fixed to be three, the growth of
$n(r,N)$ is bounded by the function $3 \times 2^{r-1}$.
If $d_H = 4$ this means that for small values of $r$ the distribution
$n(r,N)$ may not grow fast enough to display the correct
fractal structure. Only when the lattices are big enough
so that the first few  steps are negligible can the
dual lattice be used to extract $d_H$.  This constraint
on the growth is not present on the direct lattice, which
is thus better suited for extracting $d_H$.

\subsection{Coupling to matter}

To see how the point-point distributions (and $d_H$) change
as we include coupling to matter we looked at
both the Ising and 3-state Potts models coupled
to gravity.  These models are chosen because in both cases
the exact solution of the models is
known\footnote{The  
3-state Potts model coupled to gravity has just
recently been solved using matrix model techniques 
\cite{paris}.  The numerical
simulations we do here verify that the solution
is correct.  To obtain the critical coupling from
\cite{paris} one has to do some reformulation. 
This leads to  
$\beta_c^*= 1/2 \log [(45-\sqrt(45))
/(\sqrt(47)-2)].$
This is for the spins placed on triangles.
To get the coupling for spins on vertices we use
the duality transformation for the $q$-state
Potts model
$\left ( e^{2\beta_c}-1 \right ) \left (   
e^{2\beta_c^*} - 1 \right ) = q \;$\cite{wu}.};
knowing
the exact critical coupling makes the simulations
much easier.

{\small
\begin{table}
\begin{center}
\begin{tabular}{|c|c|c|c|c|}  \hline
 & \multicolumn{2}{c|}{Ising model}
 & \multicolumn{2}{c|}{3-state Potts model}   \\
 Exponent
 & Measured & Exact & Measured & Exact  \\ \hline
 $\beta/\nu d_H$  &  0.167(3)  & 1/6  &  0.199(4)  &  1/5  \\
 $\gamma/\nu d_H$ &  0.653(8)  & 2/3  &  0.608(6)  &  3/5  \\
 $1/\nu d_H$      &  0.318(12) & 1/3  &  0.382(30) &  2/5  \\  \hline
\end{tabular}
\end{center}
\caption{ Comparing critical exponents, obtained using
finite sizes scaling in $\beta_c$, to exact values,
for the Ising and 3-state Potts models coupled
to gravity}
\label{tab3}
\end{table}

}

As shown in the case of pure gravity it is preferable
to measure on the direct lattice and so we
have placed the spins on the vertices.  In that case the
critical couplings are (as we include degenerate
triangulations):
\beq{*25}
\beta_c = \frac{1}{2} \log \left [ \frac{13 + \sqrt{7}}
{14 - \sqrt{7}} \right ] \; ({\rm Ising})   
\;\;\;{\rm and} \;\;\;
\beta_c = \frac{1}{2} \log \left [ \frac{41 + \sqrt{47}}
{47 - 2 \sqrt{47}} \right ] \; ({\rm 3-state\; Potts}).
\eeq
To verify that these are indeed the correct couplings
we have performed a standard finite size scaling analysis of
some observables related to the spin models; 
the average magnetization 
      ${\cal M} \sim N^{-\beta/\nu d_H}$,
the magnetic susceptibility 
      $\chi \sim N^{\gamma/\nu d_H}$, 
and the derivative of Binders cumulant
      $\partial BC / \partial \beta \sim N^{1/\nu d_H}$. 
The measured critical exponents are shown in Table 2, together
with the exact values with which they agree very well.
The main reason is, of course, that we know $\beta_c$,
but also including
degenerate triangulations and placing the spins on vertices 
reduces finite-size effects dramatically.

\begin{figure}
\begin{center}
\subfigure{
\epsfxsize=2.6in \epsfbox{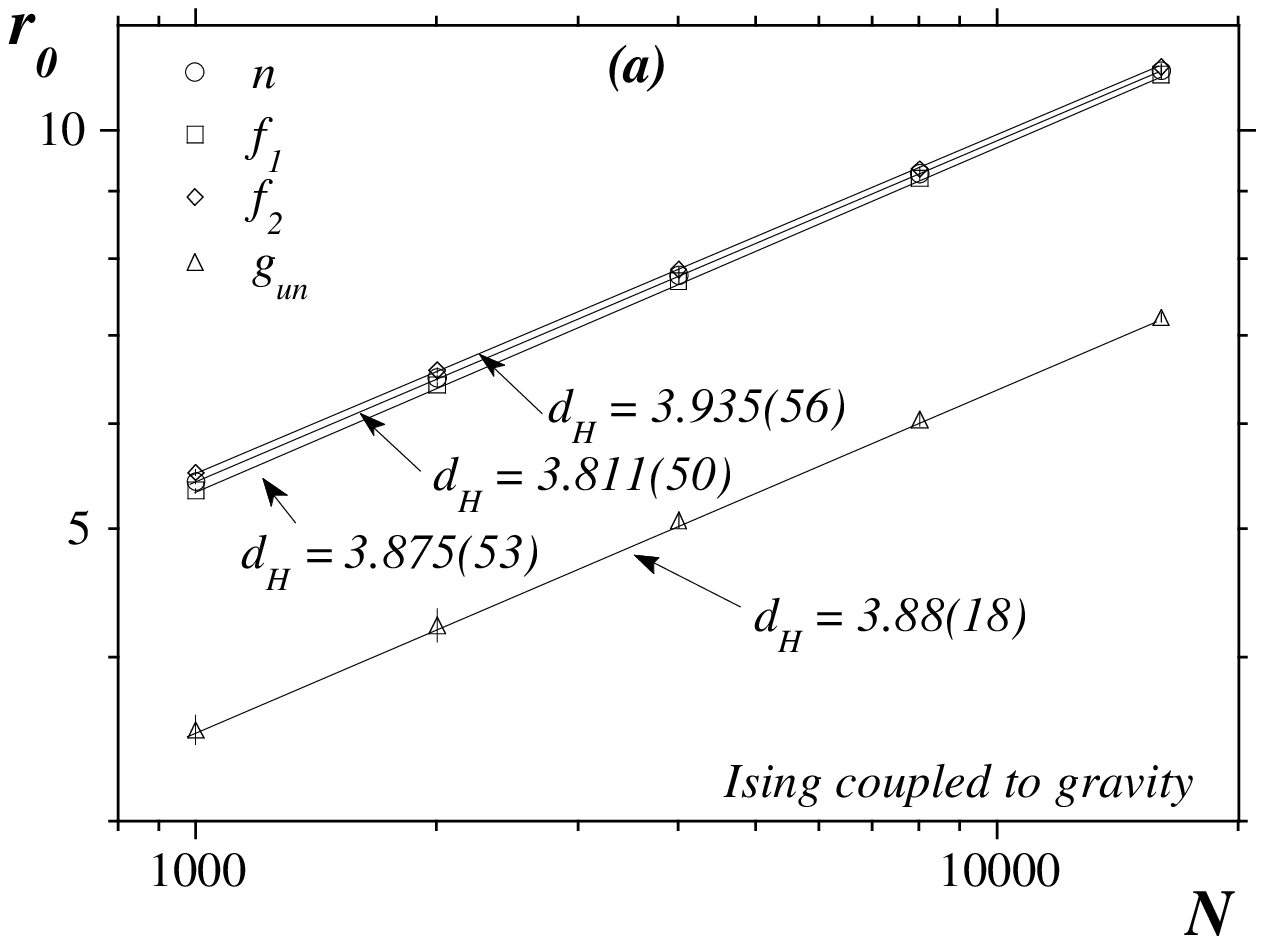} }
\subfigure{
\epsfxsize=2.6in \epsfbox{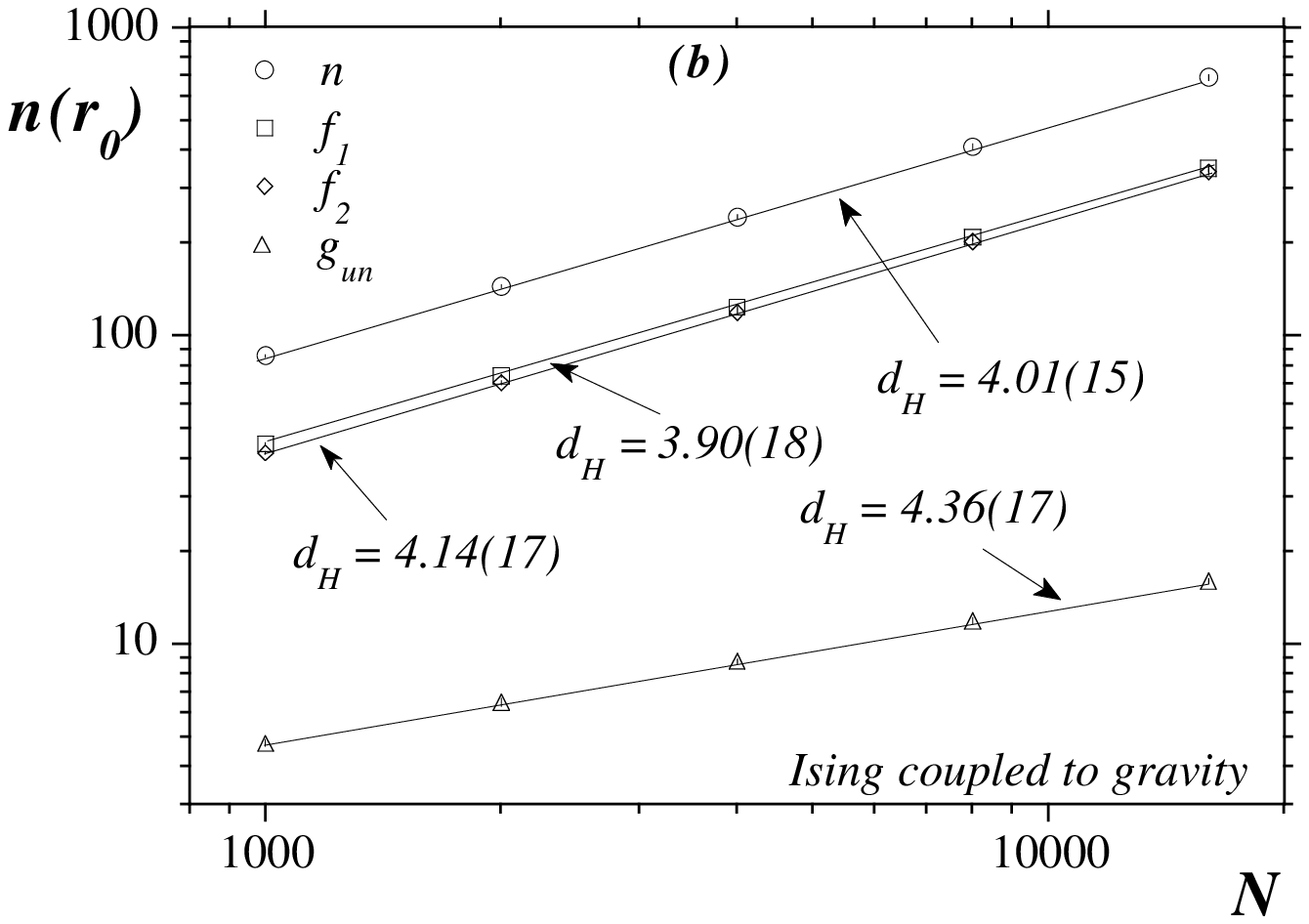} }
\end{center}
\caption
 {Volume scaling for $r_0$ and $n(r_0)$ for the 
  distributions we measured for Ising model
  coupled to gravity.  The same scaling behavior
  is used to extract $d_H$ from the slope as in 
  the case of pure gravity, except for $g_{un}(r_0)$.
  There we used $n(r_0)\sim \gamma/\nu d_H - 1/d_H$,
  substituting the exact values for $\gamma/\nu d_H$. }
\label{fig:3}
\end{figure}

Now to the distribution functions.  
The placement of spins on the vertices allows us to measure several
combinations of distributions;
$f_1(r,N)$, $f_2(r,N)$, $n(r,N)$ and 
$g_{un}(r,N)$.  We have analyzed these distributions in the
same way as for pure gravity.  In Figs.\ 3{\it a} and {\it b} we show
the scaling with volume of $r_0$ and $n(r_0)$, obtained
from fitting the distributions to the functional form 
Eq.\ (\ref{*23}).
These plots are for the Ising model but plots for the 3-state
Potts model are very similar.  The extracted Hausdorff
dimensions, for $n(r,N)$ and $g_{un}(r,N)$, are shown in
Table 3. As for pure gravity we also scaled
all the data (for $N\geq 1000$), and for pairs
of distributions, on a single curve. Resulting optimal
values of $d_H$ are listed in Table 3.  The quality of 
the scaling is shown in Figs.\ 4{\it a} and {\it b},
where we show scaling plots for  
$n(r,N)$ and $g_{un}(r,N)$ (for the Ising model).  
Again the value of $d_H$ that minimizes $\tilde{\chi}^2$
is used to scale the data.

\begin{figure}
\begin{center}
\subfigure{
\epsfxsize=2.6in \epsfbox{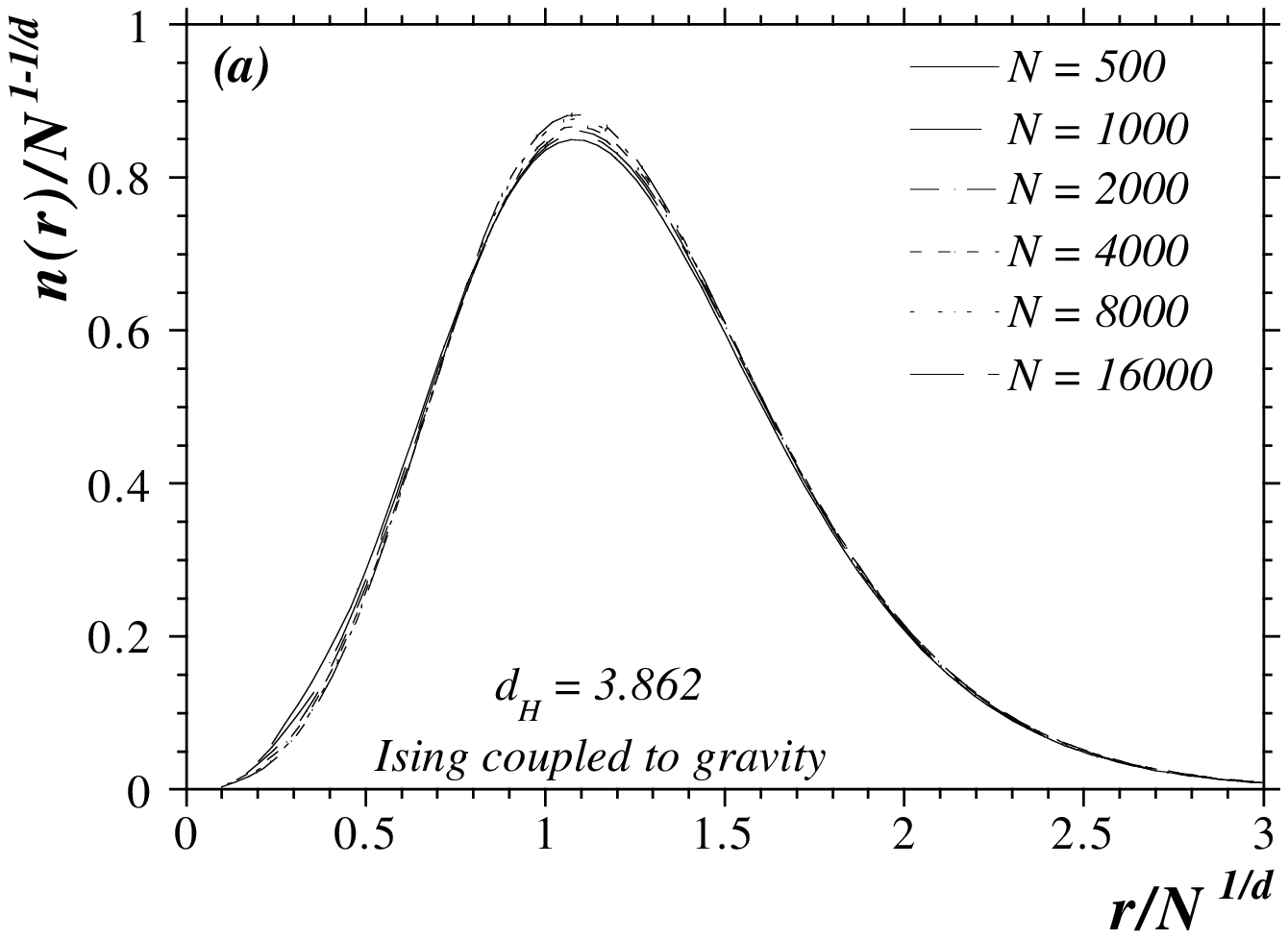} }
\subfigure{
\epsfxsize=2.6in \epsfbox{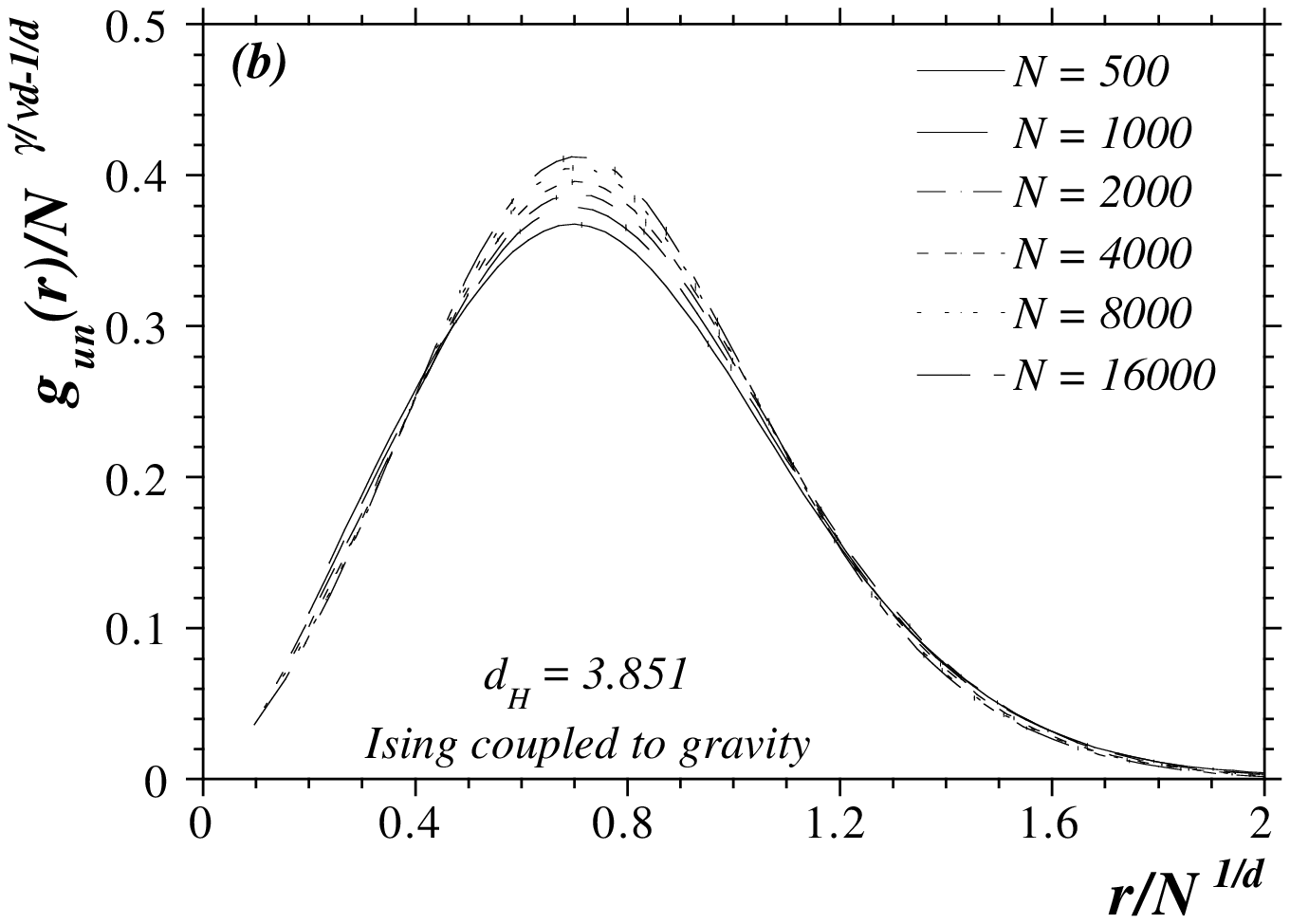} }
\end{center}
\caption
 {Collapsing the data for $n(r,N)$ and $g_{un}(r,N)$ on a single
  curve using {\it one} scaling parameter in the case
  of an Ising model coupled to gravity.}
\label{fig:4}
\end{figure}

In the case of the spin models we also measured the 
normalized spin-spin correlation function $g_n(r,N)$.  At the
critical point $g_n(r,N)$ is expected to have the following 
behavior
\beq{*37}
g_n(r,N) \sim \frac{e^{\textstyle -m(N)\;r}}{r^\eta} \;\;,
\eeq
were the mass gap $m(N)$ vanishes in the infinite
volume limit.  Surprisingly we only see
the exponential decay of the spin-spin correlator and
not the power fall-off underneath it (on a
log plot we have a straight line for some range of $r$). If we
assume that the inverse mass gap is yet another measure of
a characteristic length scale for the system, the
observed power law dependence is
an alternative measure of the Hausdorff dimension. 

Looking at the data it is clear that
the scaling hypothesis is satisfied as well here as for 
pure gravity.  What is surprising is that the extracted values of 
$d_H$, with two exceptions, are almost the same as for pure gravity.
The exceptions, for both models, are the scaling of the peak height 
of $g_{un}(r,N)$ and $d_H$ obtained from the mass gap, both
indicating larger values of $d_H$.  Why is it that
we do not seem to see any effects of the back reaction of matter on
the fractal dimension?

A possible explanation would be that 
the critical region is slightly shifted away from the
infinite volume critical coupling
at the finite volumes we simulate. This is, for example, observed
in measurements of the string susceptibility \cite{gud2},
where measured values of $\gamma_s$ peak away from $\beta_c$.
To check this we have measured $d_H$ for the Ising model over an 
interval of $\beta$.
Within errors the extracted value of $d_H$ did not change
over this interval.

In the case of pure gravity we see that the scaling of the peak 
height gives  better results.  If we believe this we get
different values for $d_H$ depending on which point-point
correlator we examine.  Looking at $n(r,N)$ we  
get $d_H \approx 3.9$ for both models, and observe
no back reaction from the matter.  The distribution 
$g_{un}(r,N)$, on the 
other hand, indicates $d_H > 4$, and indeed gives
results that might be consistent with the values
predicted in \cite{moto}.  This is supported by
the scaling of the mass gap of the spin-spin correlator.
We will return to this in the discussion section.

{\small
\begin{table}
\begin{center}
\begin{tabular}{|l|cl|cl|cl|cl|}  \hline
 & \multicolumn{4}{c|}{Ising model}
 & \multicolumn{4}{c|}{3-state Potts model}   \\
 & \multicolumn{2}{c|}{$n(r,N)$} & \multicolumn{2}{c|}{$g_{un}(r,N)$}
 & \multicolumn{2}{c|}{$n(r,N)$} & \multicolumn{2}{c|}{$g_{un}(r,N)$} \\
 & $d_H$ & $\tilde{\chi}^2$ & $d_H$ & $\tilde{\chi}^2$
 & $d_H$ & $\tilde{\chi}^2$ & $d_H$ & $\tilde{\chi}^2$     \\ \hline
{\it (a)} &&&&&&&& \\
   500-1000    & 3.758(53) &  2.6   & 3.76(12) & 0.93
               & 3.752(63) &  0.68  & 4.01(26) & 2.5    \\
   1000-2000   & 3.802(55) &  0.77  & 3.75(15) & 1.0
               & 3.787(65) &  0.29  & 4.11(18) & 1.0    \\
   2000-4000   & 3.833(56) &  1.0   & 3.73(12) & 2.5
               & 3.864(63) &  1.0   & 4.04(22) & 3.2    \\
   4000-8000   & 3.893(61) &  0.88  & 3.69(09) & 3.9
               & 3.870(73) &  0.15  & 4.11(19) & 0.41   \\
   8000-16000  & 3.870(87) &  0.35  & 3.80(10) & 0.99
               & 3.820(97) &  0.58  & 4.14(15) & 0.56   \\ \hline
{\it (b)} &&&&&&&&  \\
   1000-16000  & 3.862(74) &  1.4   & 3.851(53) & 4.5
               & 3.831(32) &  2.4   & 3.966(64) & 12.5  \\ \hline
{\it (c)} &&&&&&&&  \\
   position & \multicolumn{2}{c|}{3.875(53)}
            & \multicolumn{2}{c|}{3.88(19)}
            & \multicolumn{2}{c|}{3.879(29)}
            & \multicolumn{2}{c|}{4.141(58)}   \\
   height   & \multicolumn{2}{c|}{4.01(15)}
            & \multicolumn{2}{c|}{4.36(18)}
            & \multicolumn{2}{c|}{3.900(41)}
            & \multicolumn{2}{c|}{4.424(35))}   \\
   mass gap & \multicolumn{4}{c|}{4.51(20)}
            & \multicolumn{4}{c|}{4.56(43)} \\  \hline
\end{tabular} 
\end{center}
\caption
{Extracted values of $d_H$ for the Ising and 3-state Potts
 models coupled to gravity.  The values are obtained in the
 same way as for pure gravity (Table 1). }
\label{tab:3}
\end{table}

}

\section{Hausdorff Dimension - Analytic results}

In this section we briefly review the continuum and matrix model
derivations of the intrinsic Hausdorff dimension ($d_H$) of the 
surfaces
generated by the coupling of $2d$ gravity to matter 
\cite{wat,KKMW,moto,KN}.
There are several potentially inequivalent ways to define an
appropriate measure of the fractal dimensionality of random surfaces.
In the original paper of \cite{KN} two methods were proposed.
In the first method one determines a power-like relation between two 
gauge-invariant observables with dimensions of volume ($V$) 
and length ($L$)
respectively, with $d_H$ determined by $V \propto L^{d_H}$.
The volume is measured by the cosmological term and the length 
by the anomalous dimension of a test fermion which couples
to the gravitational field but generates no back reaction.
This yields  
\beq{*41}
d_H = 2 {\sqrt{25 -c} + \sqrt{13-c}
\over\sqrt {25 -c} +\sqrt{1-c}}\;. 
\eeq 
In the second method one considers the diffusion of a test fermion 
field
and determines $d_H$ by the short-time come-back probability 
$p(\tau) \propto \tau^{-d_H/2}$.
The authors were able to determine the Hausdorff dimension in a double
power series expansion in $\epsilon =D-2$ and ${-1 \over c}$, 
where $D$ is the classical dimensionality of 
the surface and $c$ is the 
central charge of the matter coupled to gravity.  
In \cite{moto} this second method was applied instead to a scalar 
field {--} one considers the diffusion equation for a random walk
on the ensemble of $2d$ manifolds determined by the Liouville action.
This yields
\beq{*42}
d_H = -2 {\alpha_1 \over \alpha_{-1}} 
= 2 {\sqrt{25 -c} + \sqrt{49-c}\over\sqrt{25-c} +\sqrt {1-c}}\;, 
\eeq 
where $e^{\alpha_1 \phi}$  corresponds to the cosmological constant
operator, which has dimension one, and $e^{\alpha_{-1}\phi}$ 
corresponds
to the Liouville dressing of the Laplacian, which requires it to be of
conformal dimension $-1$.

In the matrix-model/dynamical triangulation approach the transfer 
matrix 
formulation can be used to
obtain an expression for the Hausdorff dimension in the case of pure
gravity \cite{wat,KKMW}.
One finds $d_H=4$ in agreement with Eq.\ (\ref{*42}) for $c=0$.

For the case of pure gravity this result can be compared with  
\cite{KKMW}. Using matrix model results  it 
is
possible to show that 
\beq{*43}
\rho (L; D) = {3\over 7 \sqrt{\pi}} {1 \over D^2} 
\left(x^{-5/2} + {1\over 2} x^{-3/2}
+ {14\over 3}  x^{-1/2} \right) e^{-x},
\eeq
where $\rho (L;D)$ is the number of boundaries separated by geodesic 
distance $D$ from a loop of length $L$ with one marked point, and the 
scaling
variable $x={L \over D^2}$.
Now one can consider the quantities $<L^n> = \int_{a} ^{\infty}dL L^n 
\rho (L;D)$, where $a$ is the lattice constant. From Eq.\ (\ref{*43})
it can be shown that:
\begin{eqnarray}
<L^0> &\simeq &const\times D^3 a^{-3/2} + const D a^{-1/2} + const 
D^0\\
<L^1> &\simeq &const\times D^3 a^{-1/2} + const D^2\\
<L^n> &\simeq &const\times D^{2n} \;\;\; (n \geq 2).
\end{eqnarray}
Then, using the definition $\langle L^0\rangle \propto r^{d_H-1}$,
one can read off the Hausdorff dimension $d_H =4$, which agrees
with the continuum result and our numerical results based on scaling  
arguments. 
This result is not universal because of the explicit lattice 
dependence
in $<L^0>$. One obtains the same result, however, from the second and 
higher moments provided one assumes that $<L^2>$ scales like the area 
$A$.
The result thus appears to be universal.

The general situation is, however, far from clear. 
One case where there is an obvious discrepancy seems to be the 
$(2,2k-1)$ series of minimal models coupled to gravity. It is 
possible to
extend the continuum Liouville theory analysis to these models after 
taking
into account the fact that these non-unitary models possess
operators in the matter sector with negative conformal dimensions. 
It is also possible to use the results obtained in \cite{GK}
to calculate the Hausdorff dimensions for models (with `$k$' even).
We find that the results thus
obtained do not agree with each other except for the cases $k=1,2$.

The expression for the distribution of loops at a geodesic distance
`D' for the $(2,2k-1)$ models coupled to gravity (for even `$k$')  was
computed in \cite {GK}. They find that 
\beq{*46}
\rho(L;D) \simeq {1\over D^{1\over \sigma}} 
\left[{\gamma_1 \over \gamma_2 \Gamma(\sigma)} x^{-\sigma -2}(2\sigma 
+ 1
+ x ) + {x^{\sigma} \over \Gamma(\sigma + 1)} \right] e^{-x},
\eeq
where $\sigma = k-3/2 $ and $\gamma_1 , \gamma_2$ are `$k$' dependent
constants.
Using the same arguments as in the case of pure gravity we can 
compute 
$d_H = (2k-1)/(2k -3) +1$.

The continuum result of Kawamoto can also be extended to this case, 
with
the difference being that the cosmological constant is not the 
dressing of
the identity operator but  of the operator with the lowest conformal
dimension. Similarly the dressing condition for the Laplacian is that
the Liouville field has dimension $-1-\Delta_{min}$.
 
Then one obtains:
\begin {eqnarray}
\alpha_+ &=& {-k \over \sqrt{ 2k-1}} \\
\alpha_{-1-\Delta_{\rm min}} &=& {-1 \over \sqrt{ 2k-1}}\left( 2k + 1 -
\sqrt{ 32k -15}\right) \\
d_H&=& {4k \over -2k -1 +  \sqrt{ 32k -15}}\;\;\;\;.
\end{eqnarray}
It is possible to replace the dressing of the Laplacian with the 
condition
that the dressing of the Laplacian involves the identity operator and 
not
the minimal dimension operator, in which case we obtain:

\beq{*410}
d_H  = {8 \over -2k -1 +  \sqrt{(4k^2 + 20k -7)} }\;\;.
\eeq

Thus for this class of models we find an obvious discrepancy between
the matrix model and the continuum formulations.
These models are not, unfortunately, amenable to numerical 
simulations 
to resolve this disagreement.

\section{Discussion} 

We have studied a class of correlation functions defined along
geodesic paths in the dynamical triangulation formulation of
two-dimensional gravity. The critical nature of this theory is
revealed in the observation that these correlators satisfy
a scaling property. The origin of this scaling behavior
can be attributed to the existence of a dynamically
generated length scale in two-dimensional gravity.
Furthermore the power relation between this linear scale and
the total volume allows us to extract a fractal dimension
characterizing the typical quantum geometry. For pure gravity
we estimate $d_H=3.83(5)$, which is close to the analytic
prediction $d_H=4$. Our numerical method constitutes by far
the most reliable method yet investigated for extracting this
fractal (Hausdorff) dimension.

Encouraged by this result we have studied two simple spin
models coupled to quantum gravity $-$ 
the Ising  and $3$-state Potts models. 
As we have indicated there are no truly reliable analytic predictions
concerning the nature of the fractal geometry
for these values of the matter central charge.
The inclusion of matter fields allows us to define two
independent correlation functions which we have 
termed $f_1$ and $f_2$. The usual geometrical correlator
counting the number of sites at geodesic distance $r$ is just
the sum $f_1+f_2$, whilst the weighted difference $f_1-\frac{1}{q-1}f_2$
yields the (unnormalized) spin correlator.

For both types of correlation function in either the Ising or
3-state Potts cases we see good evidence for scaling.
From the geometrical correlators the Hausdorff
dimension we extract is statistically consistent with
its value for pure gravity. Taken at face
value this would seem to
indicate that the back-reaction of the critical spin system on
the geometry is insufficiently strong to alter the
Hausdorff dimension for these values of the central charge. 
This is supported by our best overall scaling fits to
the spin correlator, which yield comparable values for $d_H$.

The picture is somewhat different if 
we use only the scaling of the peak 
height to
estimate a value for $d_H$ $-$  now
a shift in $d_H$ is observed to values somewhat above four.
Indeed these estimates for $d_H$ 
are not inconsistent with the predictions of the formula
derived in \cite{moto}. Since the peak scaling
appears to suffer from smaller finite size
effects than other quantities in the case of pure gravity 
(it gives $d_H=4.040(98)$) it is
possible that it is also a more reliable channel in which to look
for signs of back-reaction in the case of spin models. These
estimates for $d_H$ are also favored by examining the scaling of
the spin correlation length extracted from the normalized 
correlation function. Without good theoretical
reasons for believing in such a favored channel, however, it is
probably more sensible to ascribe the differences in
our estimates for $d_H$ to the presence of rather
large scaling violations at these lattice sizes. 
 
One alternative scenario might be that the observed effects
are due to the presence of two linear scales; the
geometrical scale and another characterizing the critical
spin correlations. Thus two fractal dimensions might
be possible; one the (true) Hausdorff dimension associated
with the geometry, and another revealed only in the spin
channel. 
We can see how these two scales could coexist
by considering the
correlation functions $f_1$ and $f_2$. We have seen that
both of these quantities appear to scale in a similar
fashion. Indeed the exponents associated with the overall
volume prefactors are the same. Yet the weighted difference
of $f_1$ and $f_2$, $f_1-\frac{1}{q-1}f_2$, 
yields the spin correlator which is
constrained to have a different dependence on the volume.
This implies that both $f_1$ and $\frac{1}{q-1}f_2$ are composed of
identical leading terms together with subdominant terms.
The simplest scenario for $f_1$ might be
\begin{equation}
f_1\left(r,N\right)=N^x\phi\left(\mu\right)+N^y\psi\left(\mu^\prime\right)
\qquad (x>y)  \;\;.
\end{equation}
A similar expression would hold for $f_2$. 
The
idea is that the distribution $n(r,N)$ is
determined by the lead terms whilst for the
precise linear combination making up $g_{un}(r,N)$
this piece cancels and we are left with
the subdominant piece $\psi\left(\mu^\prime\right)$.

The exponent $x$ is just $1-1/d_H$ and the scaling variable
$\mu=r/N^{1\over d_H}=r/l_G$ measures the geodesic distance 
in units of the induced geometrical scale $l_G$. Similarly
the exponent $y$ is related to the spin susceptibility exponent
$y={\gamma\over\nu d_S}-{1\over d_S}$
and $\mu^\prime=r/N^{1\over d_S}=r/l_S$ is a possible new scaling variable
associated with the spin scale $l_S$. In flat
space the critical spin scale $l_S$ is just identified with
the linear scale (here $l_G$) and $d_S=d_H$. It is not
clear on a dynamical lattice that this is necessarily so;
one could imagine a scenario in which the geometrical
scale varies anomalously with the spin scale $l_G\sim
\left(l_S\right)^\omega$. The quantity $\omega$ would then
constitute a new exponent characterizing the coupled matter-gravity
system.

If this scenario were to be
realized then the numerical estimates of
these exponents would favor a situation in which
the spin correlation length diverged {\it more slowly} with
volume than the gravitational (geometrical) scale. This
might serve as a partial explanation of the observed
exponential behavior of the (normalized) spin correlator
at the critical point - unlike
flat space critical models the correlation length in
a dynamical lattice can never
reach the typical linear size of the lattice.

In the absence of any explicit transfer matrix type
solutions for these unitary minimal models it would seem that
further high resolution numerical work will be needed to
resolve these important issues.

\vspace{4mm}
{\bf Acknowledgements}

This research was supported by the Department of Energy, USA,
under contract No. DE-FG02-85ER40237 and by research funds
from Syracuse University.  We would also like to acknowledge
the use of NPAC computational facilities and the valuable 
help of Marco Falcioni.

\end{document}